\documentclass[aps,prb,groupedaddress,showpacs,amsmath,amssymb,twocolumn]{revtex4}
\usepackage{graphicx}
\usepackage{amsmath}
\begin{document}

\title{\textbf{Transition from one-dimensional antiferromagnetism to three-dimensional antiferromagnetic order in single-crystalline CuSb$_{2}$O$_{6}$}}
\author{A. Rebello}
\affiliation{Physics Department, P. O. Box 173840, Montana State
University, Bozeman, MT 59717-3840, USA}
\author{M. G. Smith}
\affiliation{Physics Department, P. O. Box 173840, Montana State
University, Bozeman, MT 59717-3840, USA}
\author{J. J. Neumeier}
\affiliation{Physics Department, P. O. Box 173840, Montana State
University, Bozeman, MT 59717-3840, USA}
\author{B. D. White}
\affiliation{Physics Department, P. O. Box 173840, Montana State
University, Bozeman, MT 59717-3840, USA}
\author{Yi-Kuo Yu}
\affiliation{National Center for
 Biotechnology Information, 8600 Rockville Pike, Bethesda, MD 20894, USA}

\date{\today}%
\begin{abstract}
Measurements of magnetic susceptibility, heat capacity and thermal expansion are reported for single crystalline CuSb$_{2}$O$_{6}$ in the temperature range $5<T<350$ K. The magnetic susceptibility exhibits a broad peak centered near 60 K that is typical of one-dimensional antiferromagnetic compounds. Long-range antiferromagnetic order at $T_N$ = 8.7 K is accompanied by an energy gap ($\Delta$ = 17.48(6) K). This transition represents a crossover from one- to three-dimensional antiferromagnetic behavior. Both heat capacity and the thermal expansion coefficients exhibit distinct jumps at $T_N$, which are similar to those observed at the normal-superconducting phase transition in a superconductor. This behavior is quite unusual, and is presumably associated with a Spin-Peierls transition occurring as a result of three-dimensional phonons coupling with {\it Jordan-Wigner-transformed} Fermions. \end{abstract}

\maketitle


\section{INTRODUCTION}

Although a clear physical picture for three dimensional antiferromagnets has emerged, the understanding of one dimensional (1D) antiferromagnets (AFMs) is less clear.\cite{affleck} In these Heisenberg chain systems, quantum effects play an important role, but long-range order is not expected.\cite{affleck,bonn} One of the interesting effects that occurs in 1D AFMs is a dimerization of the spins, known as the spin-Peierls transition,\cite{hase,pouget} which leads to a decrease in the magnetic energy without long-range magnetic order as reported in CuGeO$_3$. However, the spin-Peierls transition can also be caused by three-dimensional phonons; in this case 3D antiferromagnetic order can occur.\cite{pytte, jacobs} 

CuSb$_{2}$O$_6$ is a green-colored\cite{bystrom} insulator which orders antiferromagnetically at N\'{e}el temperature $T_N$= 8.7 K.\cite{nakua,nakua2,gibs,kato,pro} It possesses a nearly octahedral local environment of the Cu$^{2+}$ ions in contrast to other cuprates, which have a quasiplanar fourfold Cu-O coordination. \cite{muller} It undergoes a phase transition from an ideal trirutile tetragonal structure ($\alpha$-CuSb$_{2}$O$_6$) above 380 K to monoclinic ($\beta$-CuSb$_{2}$O$_6$) below while largely maintaining the quasioctahedral environment.\cite{giere} The distortion of CuSb$_{2}$O$_6$ to lower symmetry is in sharp contrast to the related trirutiles, $M$Sb$_{2}$O$_6$, which remain tetragonal ($M$ = Co and Ni) and order antiferromagnetically at 12.64 K and 2.5 K, respectively. \cite{donaldson, reimers, ramos1, ramos2, ehrenberg, nakua} 

The magnetic sublattice of CuSb$_{2}$O$_6$ has been studied previously.\cite{nakua, nakua2,gibs,kato} Two investigations\cite{nakua,gibs} noted weak magnetic superlattice peaks below $T_N$. The first was unable to refine the structure using Rietveld analysis. Two possible magnetic structures were proposed and the authors established\cite{nakua2} that the magnetic moment on the copper site was 0.5$\mu_B$. The second investigation also proposed two magnetic structural models, and ultimately found only one to be consistent with magnetic susceptibility data.\cite{gibs} The single-crystal investigation by Kato et al. presents the highest-quality data, and their magnetic structure model appears to deserve the most confidence,\cite{kato} although some magnetic reflections could not be indexed. The suggested model involves Cu moments ordered ferromagnetically along $b$ with a magnetic wave vector of ($\pi$/a,0,$\pi$/c).

The 1D model for the magnetic susceptibility of linear chain spin-$1/2$ systems due to Bonner and Fisher agrees well with the magnetic susceptibility\cite{nakua2, gibs} of CuSb$_{2}$O$_6$. The interchain-to-intrachain coupling ratio was estimated\cite{nakua} at 2$\times$10$^{-3}$. This is about 10 times smaller than reported\cite{hase2} for CuGeO$_3$. Electronic structure calculations for CuSb$_{2}$O$_6$ reveal an unusual quasi-1D magnetic ground state driven by orbital ordering, which is attributed to the presence of competing in- and out-of-plaquette orbitals and strong electronic correlations. The 1D superexchange interaction (Cu-O-O-Cu) is along the [110] direction, via the apical oxygen ions, rather than the in-plaquette oxygens. This unique orbital exchange is believed\cite{kasi} to be at the core of the unusual 1D magnetism in CuSb$_2$O$_6$.

In this report, measurements of the magnetic susceptibility, heat capacity, and thermal expansion of CuSb$_{2}$O$_6$ are reported. A broad peak at $\sim$ 60 K in the magnetic susceptibility indicates the presence of one-dimensional antiferromagnetism and three-dimensional antiferromagnetism appears at $T_N$ = 8.7 K. Subtraction of the heat capacity for the non-magnetic analog ZnSb$_{2}$O$_6$ allows determination of the magnetic entropy, which clearly shows that the one-dimensional antiferromagnetism begins to appear at $\sim$ 115 K. Both the heat capacity and thermal expansion coefficients reveal distinct jumps at $T_N$, similar to what is typically observed at a normal-superconductor phase transition. The heat capacity data reveals an energy gap below $T_{N}$ of magnitude $\Delta$ = 17.48(6) K. The unusual character of this phase transition suggests that it is a  crossover between one-dimensional antiferromagnetism and three-dimensional antiferromagnetic order that is associated with a Spin-Peierls transition occurring as a result of three-dimensional phonons coupling with {\it Jordan-Wigner-transformed} Fermions. The thermal expansion reveals extremely unusual and anisotropic behavior attributed to anharmonic lattice vibrations resulting from the presence of short-range antiferromagnetic order. The thermal expansion results are compared with published\cite{hei} x-ray data for CuSb$_{2}$O$_6$, demonstrating excellent agreement. Spin-flop transitions for magnetic fields applied along both the $a$ and $b$ axes are observed.

\section{experimental}

Single crystals used in this work were grown by chemical vapor transport, as described elsewhere.~\cite{pro} Two CuSb$_{2}$O$_6$ crystals were characterized by magnetic and heat capacity measurements using a Quantum Design Physical Property Measurement System and they yielded identical results. The orientation of the single crystals was determined using Laue x-ray diffraction. We observed twinning in the CuSb$_{2}$O$_6$ crystals. In most cases, the Laue diffraction on a sample selected for orientation confirmed the presence of two crystals (primary and secondary) with the same $b$ axis. The [100] and [001] directions of the primary crystal coincided with the [013] and [101] directions of the secondary crystal. The sample was then polished to eliminate the secondary crystal and therafter could be oriented along the principle crystallographic axes for measurements. The final Laue images of these polished samples revealed no twinning. 

A polycrystalline sample of ZnSb$_{2}$O$_6$ was prepared\cite{katsui} for heat capacity measurements by reacting well-mixed, stoichiometric amounts, of Sb$_2$O$_3$ and ZnO$_2$ in air at $600\,^{\circ}\mathrm{C}$ for 12 h. The powder was reground, pelletized, and sintered at $800\,^{\circ}\mathrm{C}$ in air for 12 h. X-ray diffraction revealed single-phase nature with lattice parameters $a$ = 4.66 \AA\ and $c$ = 9.24 \AA, in agreement with a prior report.\cite{mats} ZnSb$_{2}$O$_6$ single crystals were grown using chemical vapor transport. Polycrystalline ZnSb$_{2}$O$_6$ powder was sealed in an evacuated quartz tube and the hot and cold ends of the tube were kept at 800 and 750 $^{\circ}$C, respectively for 100 h. Well-faceted crystals were obtained, which exhibited good Laue imaged with no twinning. 

Thermal expansion measurements were performed on a CuSb$_{2}$O$_6$ sample with dimensions of about $1.27$, $1.84$, and $0.98$ $mm$ ($a$, $b$, and $c$ axes, respectively). Measurements of a second sample confirmed the results. Thermal expansion was also measured on an oriented ZnSb$_{2}$O$_6$ single crystal with dimensions 1.85, 1.22 and 0.38 mm ($a$, $b$, and $c$ axes, respectively). The quartz dilatometer\cite{John} used for the thermal expansion measurements has a sensitivity to changes in length of 0.1 \AA. Measurements along each axis consist of 1700 data points and were repeated 2-3 times and averaged. The data are corrected for the empty-cell effect and for the relative expansion between the cell and the sample. Note that the relative resolution of our dilatometer cell is about 4 orders of magnitude higher than that possible with x-ray or neutron diffraction.

\section{results}

\begin{figure}[lt]
\includegraphics[width=8cm]{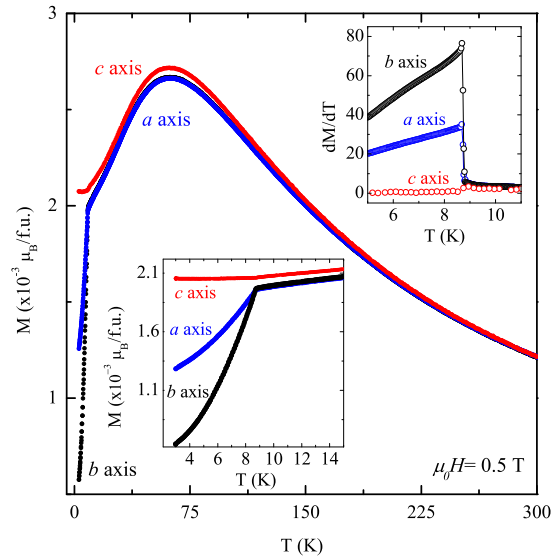}
\caption{\label{mvst} (color online) (a) Temperature (\emph{T}) dependence of
magnetization ($M$) when the magnetic field is applied along the $a$, $b$ and $c$ axes. The lower inset shows $M(T)$ expanded near $T_N$. The upper inset shows the $T$ dependence of derivative of magnetization near $T_N$.}
\end{figure}

The main panel of Figure \ref{mvst} shows the temperature (\emph{T}) dependence of the magnetization ($M$) with the magnetic field ($H$) applied along each of the principal crystallographic axes. While cooling from 300 K, $M$ increases with decreasing $T$ and exhibits a broad maxima near 60 K. This behavior is typical of systems consisting of 1D chains of magnetic ions. Accordingly, we have used the Bonner-Fisher model \cite{bonn} for the spin 1/2 AFM Heisenberg chain to compare to the observed behavior with good agreement. The fits yield an exchange coupling strength $J$ = 47.2(2) K, g$_{a}$= 2.20(1), g$_{b}$= 2.20(1) and g$_{c}$= 2.24(2), which indicates a strong spin-orbit interaction and, perhaps, some minor g-factor anisotropy. Anisotropy in the g-factor was observed through electron-spin resonance measurements.\cite{hei} At $T_N$ = 8.7 K long-range antiferromagnetic (AFM) order is observed, as reported previously. \cite{kasi,gibs} Thus, the transition at $T_N$ represents a crossover from one- to three-dimensional antiferromagnetic behavior

Prominent anisotropy is observed below $T_N$ in $M$, as shown in the lower inset of Figure \ref{mvst}. This anisotropy is clearly evident in $dM/dT$ versus $T$ for the three field directions; $dM/dT$ for $H_{a}$ and $H_{b}$ exhibits an abrupt increase at $T_{N}$ at the AFM transition while cooling whereas it exhibits a small decrease for $H_{c}$, as seen in the upper inset of Fig. \ref{mvst}; the subscript $i$ in the symbol $H_i$ denotes the direction along which $H$ was applied. This behavior is typical for AFM systems, and is largely consistent with the proposed ordering of the magnetic moments.\cite{gibs,kato} 

\begin{figure}[tr]
\includegraphics[width=8cm]{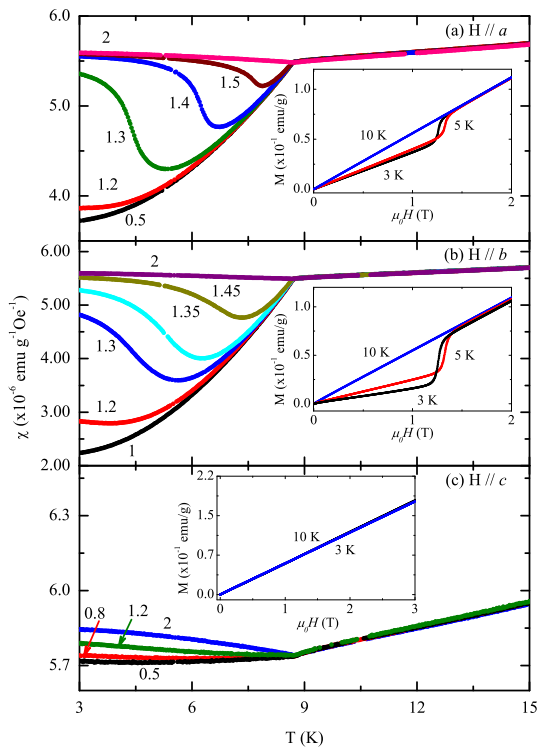}
\caption{\label{mvstdiffH} (color online) Temperature dependence of magnetization at different magnetic fields applied along the (a) $a$ (b) $b$ and (c) $c$ axes. The numbers indicate the magnetic field strength($\mu_{0}H$) in T. The insets show the field dependence of $M$ for $H$ applied along the respective axes. }
\end{figure}

To further examine the AFM ordered state, the temperature dependence of $M$ at different magnetic fields applied along each crystallographic axis was measured. Fig. \ref{mvstdiffH} shows $M(T)$ for magnetic field along each axis. For field strengths 0.5 and 1.2 T along $a$, $M(T)$ exhibits a downward drop below $T_{N}$= 8.7 K [Fig. \ref{mvstdiffH}(a)]. However, for 1.3 T, this downward drop is followed by an increase below 5 K while cooling, which could be attributed to a spin-flop transition where the spins turn from a parallel orientation to a perpendicular orientation, thereby lowering their energy. Furthermore, the spin-flop transition shifts towards higher temperatures with increasing field strengths. Note that the spin-flop transition is also manifested as abrupt jumps in $M(H)$, for temperatures below $T_{N}$, as shown in the inset of Fig. \ref{mvstdiffH}(a); in these data, it is evident that the transition occurs at lower field for lower temperatures. This is due to the lowering of total energy of the spin-flop configuration compared to the antiferromagnetic state with increasing anisotropy and coupling strength as temperature decreases. Interestingly, $M(T)$ and $M(H)$ when $H$ is applied along $b$ [Fig. \ref{mvstdiffH}(b)] exhibit similar behavior, indicating that the spin-flop transition occurs in CuSb$_{2}$O$_6$ for $H$ along both the $a$ and $b$ axes. Since Heinrich et al. [\onlinecite{hei}] observed a spin-flop transition for the $b$-axis and not for directions perpendicular to $b$, it was suggested that the easy antiferromagnetic axis lies along the $b$ direction. However, our results confirm that CuSb$_{2}$O$_6$ orders below 8.7 K with easy antiferromagnetic axis along either the $a$ or $b$ directions, in agreement with a prior report,\cite{gibs} which may reflect the presence of significant magnetic disorder, or weak differences in the magnetocrystalline energy for these two directions. As expected, we do not observe any spin flop transition in $M(T)$ and $M(H)$ for $H$ along $c$ [Fig. \ref{mvstdiffH}(c)].

\begin{figure}[rt]
\includegraphics[width=8cm]{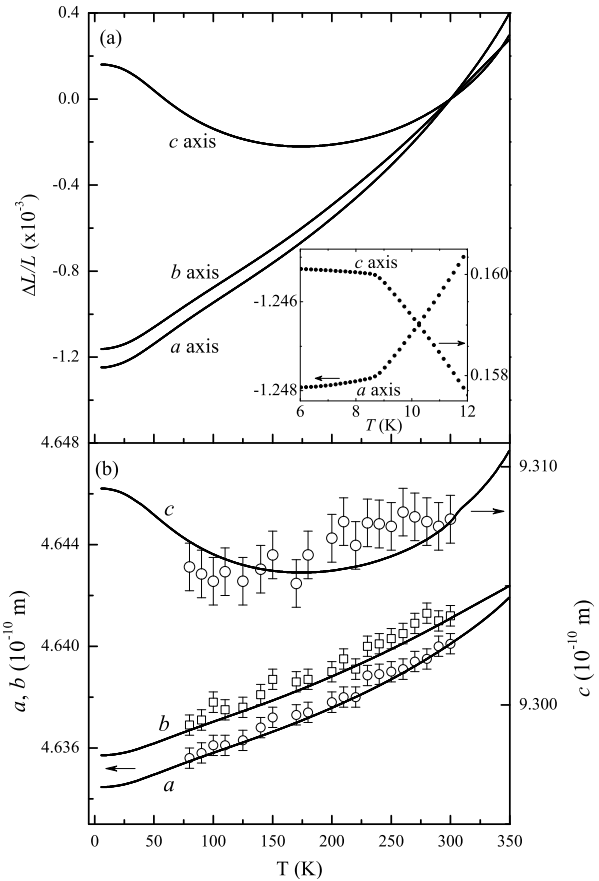}
\caption{\label{ltevst} (a) Temperature (\emph{T}) dependence of the linear thermal expansion ($\triangle$$L/L_{300}$) along the $a$, $b$,
and $c$ axes. Each curve contains more than 1700 data points. Inset shows the expanded region around $T_{N}$. (b) Comparison of temperature dependence of lattice parameters using dilatometer (solid lines, this work) and x-ray diffraction (open symbols, Ref. \onlinecite{hei}) techniques.}
\end{figure}

In Figure \ref{ltevst}(a), the linear thermal expansion normalized to the length at 300 K, $\triangle$$L/L_{300}$, is plotted for the $a$, $b$ and $c$ axes. Here, $L_{300}$ is the length of the sample at $T=$ 300 K. Upon warming from 5 K, $\triangle$$L/L_{300}$ along $a$ and $b$ increases. The magnitude of $\triangle$$L/L_{300}$ along b shows a variation of $1.16(2)\times10^{-3}$ between 5 and 300 K. On the other hand, $\triangle$$L/L_{300}$ along $c$ decreases while warming from 5-150 K and thereafter increases with further increasing temperature. The inset of Fig. \ref{ltevst}(a) displays $\triangle$$L/L_{300}$ along $a$ and $c$ on the left and right scale, respectively, for $T$ near $T_N$. Note that a clear change of slope is visible at $T_{N}$ along $a$ (similar behavior occurs along $b$) and $c$, as expected for a continuous phase transition. 

The thermal expansion data are compared to data obtained using x-ray diffraction \cite{hei} (open symbols) in Fig. \ref{ltevst}(b). Our $\triangle$$L/L_{300}$ data were multiplied by the respective lattice parameter at room temperature, and the result was then added to that lattice parameter in order to obtain the temperature dependence which appears as {\it solid lines} in the figure, which are actually about 1700 discrete data points. The lattice parameters $a$ and $b$ tend to merge near 350 K due to the well-known monoclinic to tetragonal structural transition\cite{giere} at 380 K. This comparison reveals the excellent agreement with the data obtained from diffraction as well as the outstanding resolution of our data.

\begin{figure}[lt]
\includegraphics[width=8cm]{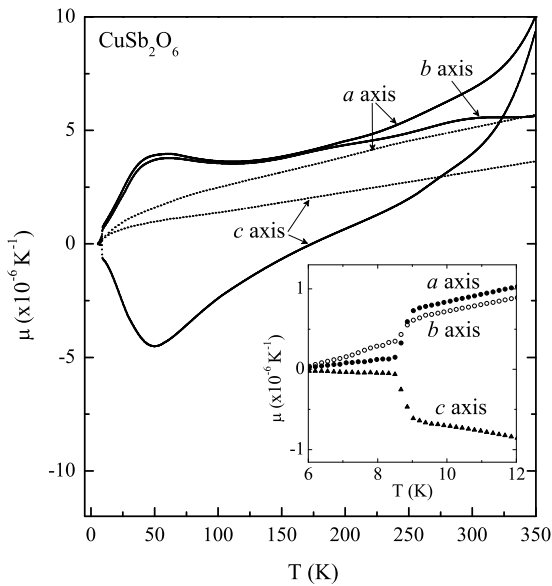}
\caption{\label{ltecvst} (a) Temperature (\emph{T}) dependence of linear thermal expansion coefficient ($\mu=(1/L_{300 K})$$\partial\triangle$$L$/$\partial$$T$) along the three crystallographic directions for CuSb$_{2}$O$_6$. Inset shows the expanded region around $T_{N}$. Dotted lines are $\mu$ for the non-magnetic analog compound ZnSb$_{2}$O$_6$.} \end{figure}

Fig. \ref{ltecvst} shows the linear thermal expansion coefficient $\mu=(1/L_{300 K})$$\partial\triangle$$L$/$\partial$$T$ for the $a$, $b$ and $c$ axes. While $\mu$ along $a$ and $b$ is positive over the entire temperature range, along $c$ it is negative at temperatures below 175 K and positive at higher temperatures. Moreover, $\mu$ displays a broad peak along $a$ and $b$ and a broad minimum along $c$ near 50 K. The expanded view of $\mu(T)$ near the AFM transition is given in the inset of Fig. \ref{ltecvst}. Abrupt jumps occur in $\mu$ at the AFM transition. While $\mu$ along $a$ and $b$ jumps downward at the paramagnetic-AFM transition on cooling, $\mu$ along $c$ shows an upward jump. Features in $\mu$ are expected at $T_N$ for a continuous phase transition, but the shape is unusual (see below). The main panel also shows $\mu$ for ZnSb$_{2}$O$_6$ along the $a$ and $c$ axes of the tetragonal crystal as dotted lines. Note that this non-magnetic analog compound exhibits monotonic behavior of $\mu$ and fails to exhibit $\mu <$ 0 along $c$. The unusual temperature dependence of the $\mu (T)$ data for CuSb$_{2}$O$_6$ is likely associated with its 1D antiferromagnetism and lower crystallographic symmetry, this point will be addressed below.

\begin{figure}[tr]
\includegraphics[width=8cm]{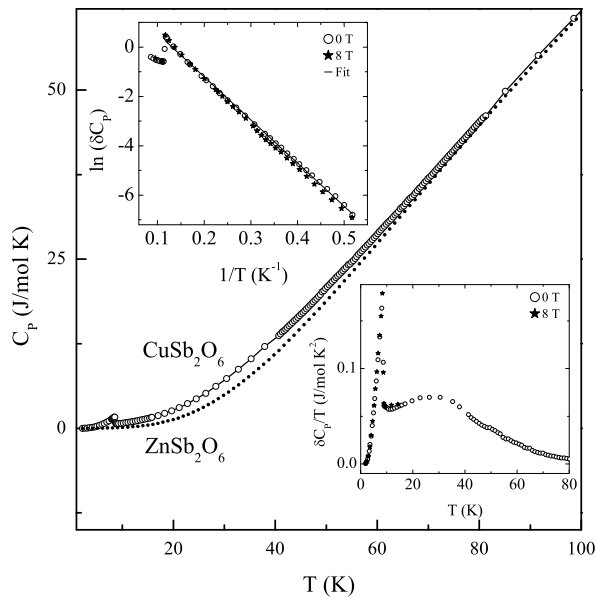}
\caption{\label{fig5} Main panel displays the molar heat capacity $C_{P}$ versus $T$ for CuSb$_{2}$O$_6$ and its non-magnetic analog ZnSb$_{2}$O$_6$ (shown as a dotted line). The upper inset shows the temperature dependence of the magnetic contribution to the heat capacity ($\delta C_{P}$) plotted as ln($\delta C_{P}$) versus 1/$T$ for 0 and 8 tesla; the data uncertainties are similar to the data point size. The solid line represents the fit to the 0 tesla data to determine the energy gap in the equation $C_P \sim A exp(-\Delta/T)$. The lower inset shows $\delta C_{P}/T$ at 0 and 8 tesla; the area under these curves is the entropy.}
\end{figure}

The heat capacity ($C_{P}$) as a function of temperature is presented for CuSb$_{2}$O$_6$ in Fig. \ref{fig5}. At the highest measurement temperature, $T$= 395 K (Data above 100 K not shown.), $C_{P}$ reaches 202.4 J/mol$\cdot$K which is close to 224.4 J/mol$\cdot$K calculated from Dulong-Petit. A prominent anomaly occurs at $T_{N}$, which will be discussed further below. Fitting to the equation $C_{P}/T = \gamma+\beta T^{2}$ over the range 9.2 K $<$ T $<$ 15.6 K yields $\gamma =$ 58.7(7) mJ/mol$\cdot$K$^2$, $\beta =$ 0.171(5) mJ/mol$\cdot$K$^4$ and a Debye temperature $\Theta_D$ = 467(5) K.

\begin{figure}[tr]
\includegraphics[width=8cm]{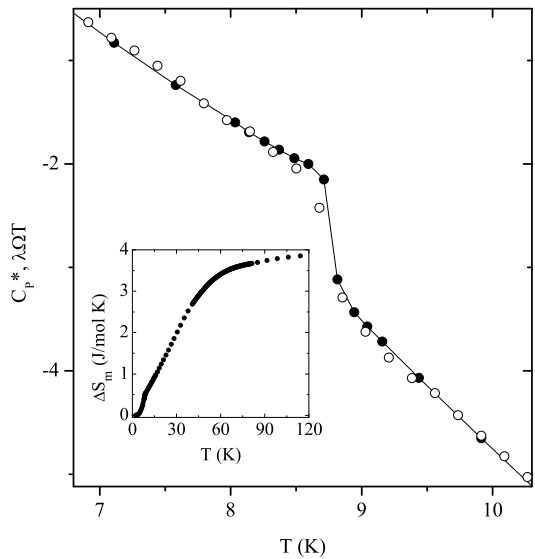}
\caption{\label{mhcvst} Molar heat capacity at zero magnetic field after subtracting the entropy contributions, $C_{P}^{*}$ (closed symbols) and $\lambda \Omega T$ versus $T$; $\lambda = -$55$\times$10$^{4}$ J/mol$\cdot$K. Inset shows $\Delta S_m$, which was obtained by integrating the $\delta C_{P}/T$ versus $T$ data over the region below 115 K.  }
\end{figure}

The main panel of Fig. \ref{fig5} also shows $C_P$ of ZnSb$_{2}$O$_6$, the non-magnetic analog of CuSb$_{2}$O$_6$. Fitting these data over the range 0.45 K $<$ T $<$ 15.6 K to the equation $C_{P}/T = \gamma + \beta T^{2}$ yields $\gamma =$ 0.02(4) mJ/mol$\cdot$K$^2$ $\beta =$ 0.11(2) mJ/mol$\cdot$K$^4$ and a Debye temperature $\Theta_D$ = 539(30) K. Subtraction of the ZnSb$_{2}$O$_6$ data from the CuSb$_{2}$O$_6$ data yields $\delta C_{P}$, which is shown in the insets of Fig. \ref{fig5} as either $\delta C_{P}$ or $\delta C_{P}/T$ in expanded views.  The broad peak in $\delta C_{P}/T$ above $T_{N}$ corresponds to the contribution of short-range-magnetic ordering to the entropy. Integrating the $\delta C_{P}/T$ versus $T$ curve (lower inset) over the region below 115 K (the temperature below which the CuSb$_{2}$O$_6$ and ZnSb$_{2}$O$_6$ data sets deviate from one another) provides the entropy change $\Delta S_m$, which is plotted in the inset of Fig. \ref{mhcvst}. An entropy change of 3.85 J/mol$\cdot$K (66.8\% of $Rln2$) is observed over this entire temperature range. If one considers only the region from just above $T_N$ to 115 K, 57.8\% $Rln2$ is obtained, so most of the entropy change occurs between 115 K and $T_N$. Note that for the region 50 K to 115 K, the contribution to $\Delta S_m$ is only 12\% Rln2. Some measurements of $C_P$ at 8 tesla, with the magnetic field parallel to the $c$ axis, were also conducted. The zero field ZnSb$_{2}$O$_6$ data were subtracted from the $C_P(T,8\, \rm{tesla})$ data to obtain $\delta C_P(T,8\, \rm{tesla})$, which is shown in the insets of Fig. \ref{fig5}. These data extend only slightly above $T_N$, where a small increase in $\delta C_P(T,8\, \rm{tesla})/T$ is evident. The  $T < T_N$ region will be discussed below.

Considering the thermodynamics of a continuous phase transition, the heat capacity in the immediate vicinity of $T_N$ can be written as\cite{sou}
\begin{equation}\label{equation1}
C_{P}=T\left(\frac{\partial S}{\partial T}\right)_{N}+\upsilon T
\Omega \left(\frac{\partial P}{\partial T}\right)_{N}. 
\end{equation}
Here, $S$, $P$, $\upsilon$ and $\Omega$ are the molar entropy, pressure, molar volume and volume thermal expansion coefficient,
respectively; the subscript $N$ signifies that Eq. (\ref{equation1}) is valid near the N\'{e}el temperature. For this analysis, it is assumed that $\Omega$ =($\mu_{a}$+$\mu_{b}$+$\mu_{c}$)$\times$sin$\alpha$, near 8.7 K, since the crystallographic angle $\alpha$ tends to saturate to 91.3$^{\circ}$ below 100 K (see Fig. 2 of Ref. [\onlinecite{hei}]). When the entropy contribution is eliminated by subtracting the term that is linear in $T$ from $C_P$, the result, $C_{P}^*$, is proportional to the product of $\Omega$ and $T$. As expected from Eq. (\ref{equation1}) and shown in Fig. \ref{mhcvst}, close to $T_{N}$, $C_{P}^*$ scales with $\Omega T$ following the relation $C_{P}^*\simeq \lambda \Omega T$. The excellent overlap of the two data sets is obtained with $\lambda = -$55$\times$10$^{4}$ J/mol$\cdot$K. Furthermore, the observation that the data sets scale suggests that the phase transition is continuous. The parameter $\lambda$ can now be used to determine $dT_{N}/dP$ since $\upsilon / \lambda = dT_{N}/dP$. The molar density at 300 K provides $\upsilon$ as 6.8 $\times$ 10$^{-5}$ m$^{3}$/mol, which then yields the pressure derivative $dT_{N}/dP$= -0.11(1) K/GPa. Note that there are no direct measurements of $dT_{N}/dP$ available thus far.

\section{discussion}

The $C_P$ data reveal a sizable $\gamma$ term of magnitude $\gamma =$ 57.4(6) mJ/mol$\cdot$K$^2$, that completely vanishes below $T_N$. Since CuSb$_{2}$O$_6$ is an electrical insulator, this term cannot be associated with the presence of conduction electrons and must have a purely magnetic origin. For a Heisenberg chain, theory\cite{johnston} reveals $\gamma$ = $(2/3)Nk_B^2/J \approx$ 120 mJ/mol$\cdot$K$^2$ using the value of $J$ obtained from fitting our magnetic susceptibility data. Thus, the obtained $\gamma$ is approximately half of the theoretical value. The decay of $C_P$ below $T_N$ is exponential, suggesting the presence of an energy gap. This gap is extracted from a linear fit to the data plotted as ln($\delta C_{P}$) versus 1/$T$ (see upper inset of Fig. \ref{fig5}), suggesting the function $\delta C_{P}$ $\sim$ $Aexp(-\Delta/T)$. The gap $\Delta$= 17.48(6) K ($\sim$1.51(1) meV) is obtained from the fit. This gap must be associated with the crossover from partially-ordered moments on non-interacting chains to three-dimensional, long-range order. Measurements of $C_P$ at an applied magnetic field of 8 T have a small influence on the obtained $\Delta$, yielding $\Delta$= 18.42(6) K ($\sim$1.59(1) meV), which suggests that the low-temperature ordered phase is more robust in non-zero magnetic field. 

The character of the phase transition at $T_N$ warrants some discussion. Normally the heat capacity and thermal expansion coefficient of a magnetic system displays $\lambda$-like character at the magnetic ordering temperature, which is associated with a divergence of measurable quantities, such as $C_P$, $\chi$, $\mu T$ and $M$ at the critical point.\cite{sou,white} The antiferromagnetic analog system CoSb$_{2}$O$_6$, for example, clearly exhibits $\lambda$-like character in $C_P$ at its AFM transition.\cite{aaron} This is not the case for CuSb$_{2}$O$_6$, where $C_P$ and $\mu$ display discontinuities at $T_N$ followed by an exponential decay in $C_P$ below $T_N$. The character of these transitions is reminiscent of that observed at the normal-superconducting phase transition, the only phase transition known to exhibit distinct jumps at the critical temperature.\cite{sou} Note that other\cite{johnston,win} 1D magnetic systems, such as CuGeO$_3$ and NaV$_2$O$_5$ exhibit $\lambda$-like character in $C_P$ and $\mu$. These systems, however, possess strong lattice distortions\cite{pouget} at the spin-Peierls transition (in the case of CuGeO$_3$) and charge-ordering transition (in NaV$_2$O$_5$). This magnetic-lattice coupling may play a role in the large $\lambda$-like features in $\mu$ at their respective transitions. In the case of CuSb$_{2}$O$_6$, it exhibits a transition from 1D to 3D order at $T_N$, \emph{without} an obvious presence of strong magnetic-lattice coupling.  

A possible scenario that agrees with the observed discontinuity of specific heat, the energy gap below $T_N$, and a weak magnetic-lattice coupling is described as follows. The Spin-Peierls transition originates from the coupling between the 1-D {\it Jordan-Wigner-transformed}  Fermions and 3-D phonons. Similar to the usual Peierls transition, it opens up an energy gap that is very much BCS-like.\cite{pytte,jacobs} In our system, there is a collection of quasi-1D spin chains that run parallel to one another. The same group of phonons that couple spins on one chain may also couple spins on the other chains. This introduces an effective inter-chain coupling. If this coupling remains marginally weak, upon the dimerization (Spin-Peierls) transition, the energy gap opened should behave essentially like the case of a single chain while the dimerization ordering of all chains will then appear as a 3-D ordering. If this is the case, the specific-heat discontinuity occurs in the same manner as the BCS case and the spin-energy gap appears as an additional contribution of specific heat the same way as the heat capacity contribution of superconducting electron pairs. This ordered state is likely an alignment of the 1D chains, rather than a true 3D ordered state; this may be the source of difficulty in identification\cite{nakua2,gibs,kato} of the magnetic lattice below $T_N$ and the disorder suggested from the spin-flop transition's occurrence along both the $a$ and $b$ axes.

The nearly tetragonal crystal structure of CuSb$_{2}$O$_6$ results in almost identical positive thermal expansion behavior observed along the $a$ and $b$ axes below 200 K. Above this temperature $\mu (T)$ differs for these two axes due to the well-known monoclinic to tetragonal structural transition at 380 K.\cite{giere} In contrast, the thermal expansion along $c$ is negative at temperatures below about 170 K. Negative thermal expansion can originate from unusual phonon modes that become active at low temperatures\cite{barerra} and are often observed in low-D systems.\cite{san,alwyn} However, consideration of $\mu (T)$ for the non-magnetic analog compound ZnSb$_{2}$O$_6$, along with the fact that 1D antiferromagnetism sets in below 115 K (see $\Delta S_m$ versus $T$ in the inset of Fig. \ref{mhcvst}) suggests that the unusual temperature dependence of $\mu (T)$ in CuSb$_{2}$O$_6$ is associated with the 1D antiferromagnetism.

The thermal expansion in solids is due to anharmonic vibrations, and in turn, anharmonic contributions to the elastic potential. This is a many-body potential, but pair potentials between neighboring atoms can play a significant role.\cite{barerra} In CuSb$_2$O$_6$ the onset of local 1D magnetic order occurs on cooling below $\sim$115 K, as determined from the heat capacity $\delta C_{P}$. Local order would strongly affect the pair potentials in those regions of the sample where the order occurs. This could lead to stretching or tilting of bonds as well as changes in bond strength in these regions. Even if the affected pair potentials are harmonic, the result can be a potential for the \emph{sample} that is anharmonic.\cite{barerra} Thus, the increase in magnitude of the thermal expansion coefficients below $\sim$ 115 K is most likely associated with changes in anharmonic contributions to the elastic potential of the sample as a result of the formation of short-range 1D antiferromagnetic order. Below $\sim$50 K the thermal expansion coefficients change behavior because a large proportion of the spins are ordered, and the anharmonic contributions reduce upon further cooling leading to a decrease in the magnitude of $\mu$. The negative thermal expansion along $c$ is in strong contrast to the behavior along $a$ and $b$, and is certainly connected to the 1D behavior between 8.7 K and 115 K and anharmonic lattice vibrations that specifically affect $\mu$ along this direction.  

Other types of transitions can also lead to the development of anharmonic contributions to the thermal expansion, such as in Li$_{0.9}$Mo$_{6}$O$_{17}$ where at 28 K a crossover in dimension has been associated with two-particle hopping and the formation of bosons.\cite{san} CuGeO$_3$ also exhibits\cite{win} a prominent correlation between thermal expansion and the formation of 1D magnetic correlations. 

\section{conclusions}

The thermodynamic and magnetic properties of CuSb$_{2}$O$_{6}$ were investigated. One-dimensional antiferromagnetic order among the spin $1/2$ Cu atoms is observed to occur below $\sim$115 K. This order has a strong impact on anharmonic lattice vibrations, and leads to a sizable change in the thermal expansion coefficients. At $T_N$ = 8.7 K, a crossover from one- to three-dimensional antiferromagnetic behavior occurs. This long-range antiferromagnetic order leads to distinct jumps in the heat capacity and thermal expansion coefficients, which are reminiscent of those observed at the normal-superconducting phase transition. The heat capacity data reveals the presence of a gap below $T_{N}$ of magnitude $\Delta $ = 17.48(6) K. The occurrence of three-dimensional antiferromagnetic order is attributed to a Spin-Peierls transition occurring as a result of three-dimensional phonons coupling with {\it Jordan-Wigner-transformed} Fermions. Spin-flop transitions for magnetic field applied in $a$ and $b$ axes below $T_{N}$ indicate that the easy AFM axis lies in both the $a$ and $b$ directions, which may reflect the presence of significant magnetic disorder, or weak differences in the magnetocrystalline energy for these two directions.

We thank Anton Vorontsov for valuable discussions. This material is based on the work supported by the National Science Foundation under Contract No. DMR-0907036.


\begin{thebibliography}{1}
\bibitem{affleck} I. Affleck, J. Phys.: Condens. Matter \textbf{1}, 3047 (1989)
\bibitem{bonn} J. C. Bonner and M. E. Fisher, Phys. Rev. \textbf{135}, A640 (1964).
\bibitem{hase}M. Hase, I. Terasaki, and K. Uchinokura, Phys. Rev. Lett. \textbf{70}, 3651 (1993).
\bibitem{pouget} J. P. Pouget, L. P. Regnault, M. Ain, B. Hennion, J. P. Renard, P. Veillet, G. Dhalenne, and A. Revcolevschi, Phys. Rev. Lett. \textbf{72}, 4037 (1994).
\bibitem{pytte} E. Pytte, Phys. Rev. B {\bf 10}, 4637 (1974).
\bibitem{jacobs} I. S. Jacobs, J. W. Bray, H. R. Hart, Jr., L. V. Interrante, J. S. Kasper, G. D. Watkins, D. E. Prober, and J. C. Bonner,  Phys. Rev. B 14, 3036 (1976).
\bibitem{bystrom}A. Bystrom, B. Hok, and B. Mason, Ark. Kemi. Mineral. Geol. B \textbf{15}, 1 (1941).
\bibitem{nakua}A. Nakua, H. Yun, J. N. Reimers, J. E. Greedan, and C. V. Stager, J. Solid State Chem. \textbf{91}, 105 (1991).\bibitem{nakua2} A. M. Nakua and J. E. Greedan, J. Solid State Chem. \textbf{118}, 1999 (1995).
\bibitem{gibs}B. J. Gibson, R. K. Kremer, A. V. Prokofiev, W. Assmus, and B. Ouladdiaf, J. Magn. Magn. Mat. \textbf{272}, 927 (2004).
\bibitem{kato} M. Kato, K. Kajimoto, K. Yoshimuar, K. Kosuge, M. Nishi, and K. Kakurai, J. Phys. Soc. Jpn. \textbf{71} Suppl., 187 (2002).
\bibitem{pro} A. V. Prokofiev, F. Ritter, W. Assmus, B. J. Gibson, and R. K. Kremer, J. Crystal Growth \textbf{247}, 457 (2003).
\bibitem{muller}H. M\"uller-Bauschbaum, Agnew.Chem. \textbf{89}, 704 (1977).
\bibitem{giere}E. -O. Giere, A. Brahimi, H. J. Deiseroth, and D. Reinen, J. Solid State Chem. \textbf{131}, 263, (1997).
\bibitem{donaldson}J. D. Donaldson, A. Kjekshus, D. G. Nicholson, and T. Rakke, Acta Chem. Scand. \textbf{29}, 803 (1975).
\bibitem{reimers}J. N. Reimers, J. E. Greedan, C. V. Stager, and R. Kremer, J. Solid State Chem., \textbf{83}, 20 (1989);
\bibitem{ramos1}E. Ramos, M. L. Veiga, F. Fern\'andez, and R. Saez-Puche, J. Solid State Chem., \textbf{91}, 113 (1991).
\bibitem{ramos2}E. Ramos, F. Fern\'andez, A. Jerez, C. Pico, J. Rodriguez-Carvajal, R. Saez-Puche, and M. L. Veiga, Mater. Res. Bull. \textbf{27}, 1041 (1992).
\bibitem{ehrenberg}H. Ehrenberg, G. Wltschek, J. Rodriguez-Carvajal, and T. Vogt, J. Magn. Magn. Mat. \textbf{184}, 111 (1998).
\bibitem{hase2} M. Hase, J. Magn. Mag. Materials \textbf{177-181}, 611 (1998).
\bibitem{kasi}D. Kasinatham, K. Koepernik, and H. Rosner, Phys. Rev. Lett. \textbf{100}, 237202 (2008).
\bibitem{hei} M. Heinrich, H.-A. Krug von Nidda, A. Krimmel, A. Loidl, R. M. Eremina, A. D. Ineev, B. I. Kochelaev, A. V. Prokofiev, and W. Assmus, Phys. Rev. B {\bf67}, 224418 (2003).
\bibitem{katsui} A. Katsui, and H. Matsushita, Phys. Stat. Sol. (A) {\bf203}, 2832 (2006).
\bibitem{mats} S. Matsushima, T. Tanizaki, H. Nakamura, M. Nonaka, and M. Arai, Chem. Lett {\bf10}, 1010 (2001).
\bibitem{John} J. J. Neumeier, R. K. Bollinger, G. E. Timmins, C. R. Lane, R. D. Krogstad, and J. Macaluso, Rev. Sci. Instrum. {\bf 79}, 033903 (2008).
\bibitem{sou}J. A. Souza, Y.-K. Yu, J. J. Neumeier, H. Terashita, and R. F. Jardim, Phys. Rev. Lett. \textbf{94}, 207209 (2005).\bibitem{johnston} D. C. Johnston, R. K. Kremer, M. Troyer, X. Wang, A. Kl\"umper, S. L. Bud'ko, A. F. Panchula, and P. C. Canfield, Phys. Rev. B \textbf{61}, 9558 (2000).\bibitem{white} B. D. White, J. A. Souza, C. Chiorescu, J. J. Neumeier, and J. L. Cohn, Phys. Rev. B \textbf{79}, 104427 (2009). \bibitem{aaron} A. Christian, M. G. Smith, A. Rebello and J. J. Neumeier, unpublished.
\bibitem{win}H. Winkelmann, E. Gamper, B. Buchner, M. Braden, A. Revcolevschi, and G. Dhalenne, Phys. Rev. B  \textbf{51}, 12884 (1995).
\bibitem{barerra} G. D. Barerra, J. A. O. Bruno, T. H. K. Barron, and N. L. Allan, J. Phys.: Condens. Matter \textbf{15}, R217 (2005)
\bibitem{san}C. A. M. dos Santos, B. D. White, Yi-Kuo Yu, J. J. Neumeier, and J. A. Souza, Phys. Rev. Lett. \textbf{98}, 266405 (2007).
\bibitem{alwyn} A. Rebello, J. J. Neumeier, Z. Gao, Y. Qi, and Y. Ma, Phys. Rev. B \textbf{86}, 104303 (2012).






\end{thebibliography}
\end{document}